\documentclass[11pt]{article}
\usepackage{amsfonts}
\usepackage{graphicx}
\usepackage{epstopdf}
\usepackage{epsfig}
\usepackage{amssymb}
\usepackage{setspace}
\usepackage{caption}
\usepackage{color}
\usepackage{amsmath}
\usepackage{float}
\usepackage{comment}
\usepackage{subcaption}

\newcommand{\be}{\begin{equation}}
\newcommand{\ee}{\end{equation}}
\newcommand{\bea}{\begin{array}}
	
	\newcommand{\eea}{\end{array}}

\numberwithin{equation}{section}
\numberwithin{figure}{section}

\begin{document}
	
	\begin{titlepage}
		\vspace{1cm}
		\begin{center}
			{\Large \bf Zipoy--Voorhees spacetime with Maxwell and dilaton fields:
				exact solution and equatorial geodesics}
		\end{center}
		\vspace{2cm}
		\begin{center}
			\renewcommand{\thefootnote}{\fnsymbol{footnote}}
			Haryanto M.\ Siahaan\footnote{haryanto.siahaan@unpar.ac.id}\\[4pt]
			Program Studi Fisika, Universitas Katolik Parahyangan,\\
			Jalan Ciumbuleuit 94, Bandung 40141, Indonesia
			\renewcommand{\thefootnote}{\arabic{footnote}}
		\end{center}
		
		\begin{abstract}
		We derive a new exact solution of four-dimensional Kaluza--Klein theory
		by applying the uplift--boost--reduction procedure to the
		Zipoy--Voorhees vacuum seed. The resulting Zipoy--Voorhees--dilaton
		spacetime is a static, axisymmetric, asymptotically flat configuration
		parametrised by a seed length scale $M$, the Zipoy--Voorhees
		deformation parameter $k$, and a boost parameter $\alpha$ that
		controls the electric charge. For $k=1$ it admits a regular
		black-hole interpretation; for $k\neq1$ it inherits the naked
		curvature singularity of the seed at $r=2M$. We extract the ADM
		mass, electric charge, and dilaton charge, finding that the dilaton
		charge obeys an algebraic constraint independent of the quadrupole
		deformation. The equatorial geodesics of neutral test particles and
		photons are studied in detail. A compact algebraic identity locates
		the exterior-photon-ring threshold at $k=1/2$ for all $\alpha$,
		generalising the known threshold of the vacuum Zipoy--Voorhees case.
		The innermost stable circular orbit is mapped over the $(k,\alpha)$
		plane, and the comparison with Schwarzschild is shown to depend
		qualitatively on whether seed-mass or ADM-mass units are used. We further
		compute the gravitational redshift, which diverges on the $r=2M$ surface
		and renders the near-singularity orbits effectively invisible, and the
		critical impact parameter setting the characteristic shadow size, which in
		ADM-mass units shrinks below the Schwarzschild value as the charge grows.
		\end{abstract}
	\end{titlepage}
	
	\onecolumn
	\bigskip
	
\section{Introduction}
\label{sec.intro}

The Kerr metric stands as one of the most celebrated exact solutions of Einstein's
field equations, describing the exterior spacetime of a rotating black hole and
serving as the canonical model for astrophysical black holes throughout the
universe. The no-hair theorem asserts that any stationary, asymptotically flat
black hole in general relativity is uniquely characterised by its mass, angular
momentum, and electric charge alone
\cite{Israel:1967wq,Carter:1971zc,Hawking:1971vc,Robinson:1975bv,Mazur:1982db}.
While this result is theoretically elegant, it raises an important empirical
question: do the compact objects we observe in nature truly conform to the Kerr
geometry, or do deviations exist that current and future observations might
reveal?

This question has gained considerable urgency in the era of gravitational-wave
astronomy and black-hole imaging. The detections by the LIGO--Virgo--KAGRA
collaboration
\cite{Abbott:2016blz,LIGOScientific:2021djp,LIGOScientific:2025gwtc4} and the
images produced by the Event Horizon Telescope
\cite{EventHorizonTelescope:2019dse,EventHorizonTelescope:2022xnr,
	EventHorizonTelescope:2022vjs} have opened entirely new observational windows
onto the strong-field regime of gravity, making it possible in principle to test
the Kerr (and, in the static limit, the Schwarzschild) hypothesis with
unprecedented precision. To exploit these opportunities one needs a sufficiently
broad class of exact solutions that smoothly deform the canonical black-hole
geometries in a controlled way, thereby providing concrete templates against
which observational data can be compared. In the present paper we work
exclusively within the static sector; the resulting solution is therefore a
testbed for non-Schwarzschild, non-rotating compact objects rather than a direct
alternative to Kerr. The astrophysically more relevant rotating extension is
left for future work.

A natural and well-studied family of static deformations is provided by the
Zipoy--Voorhees (ZV) spacetime
\cite{Zipoy:1966btu,Voorhees:1970ywo}. Originally constructed as a static,
axisymmetric vacuum solution of the Einstein equations, the ZV metric generalises
the Schwarzschild geometry through a single real deformation parameter $k$ that
encodes the quadrupole moment of the source: in the Geroch--Hansen scheme, and
in the parametrisation $k=1+q$ with $q$ the standard $q$-metric quadrupole
parameter \cite{Quevedo:2010}, the mass monopole and quadrupole take the form
\[
M_0 = kM,\qquad
M_2 = -\tfrac{1}{3}\,M^{3}\,k\,(k^{2}-1),
\]
so that $M_2$ vanishes at $k=1$ and grows cubically with the deviation from
spherical symmetry; the sign of $M_2$ then depends on whether $k\gtrless 1$,
with the precise correspondence between this sign and the ``prolate'' or
``oblate'' character of the source determined by the convention in which
multipoles are computed (different references in the ZV literature use
opposite conventions, see e.g.\ \cite{Quevedo:2010,Toshmatov:2019qih}).
We stress that $k$ is not a new parameter: it is the same Zipoy--Voorhees
deformation denoted $\gamma$ or $\delta$ in the original
literature \cite{Zipoy:1966btu,Voorhees:1970ywo,Kodama:2003ch} and $1+q$ in
the $q$-metric (or $\gamma$-metric) formulation \cite{Quevedo:2010}, the
dictionary being simply $k\equiv\gamma\equiv\delta=1+q$. We retain the symbol
$k$ throughout only because it keeps the exponents in the boosted line
element~\eqref{eq.ZVKKmetric} compact; readers more familiar with any of the
standard conventions may substitute accordingly. For
$k=1$ the solution reduces exactly to Schwarzschild, while $k\neq1$ yields a
geometry that departs from spherical symmetry in a precise and analytically
tractable manner. Owing to these properties, the ZV family has
attracted sustained attention as a testbed for the no-hair theorem and as a
model for compact objects with nontrivial multipolar
structure \cite{Quevedo:2010,Johannsen:2010,Bambi:2011,Arrieta:2021}.
Geodesic motion in the ZV spacetime has been studied in detail from multiple
perspectives: the structure of timelike and null orbits and the qualitative
changes induced by $k\neq1$
\cite{Herrera:1998rj,Benavides-Gallego:2018htf,Toshmatov:2019qih},
spinning test particles governed by the Mathisson--Papapetrou--Dixon equations
\cite{Toshmatov:2019bda}, and extreme-mass-ratio inspirals and their
gravitational-wave signatures, where the ZV geometry acts as a black-hole
mimicker \cite{Destounis:2023gpw}. A substantial body of work by Quevedo and
collaborators has developed the geodesic and observational phenomenology of
the ZV/$q$-metric in particular, including the motion of test particles in
the quadrupolar naked-singularity field \cite{Boshkayev:2016},
neutrino oscillations \cite{Boshkayev:2020}, the luminosity of accretion
discs around quadrupolar compact objects \cite{Boshkayev:2021lum},
gravitational lensing \cite{Boshkayev:2024lens}, the gravitational capture
cross-section \cite{Momynov:2025}, geodesic deviation \cite{Idrissov:2025},
and the constraints on quadrupole deformations from relativistic
effects \cite{Utepova:2025}. Extensions of the ZV spacetime to include
electromagnetic fields have also received considerable attention: charged
generalisations have been derived in the context of string theory
\cite{Yunusov:2025chw}, and the magnetised generalisation obtained via the
Harrison transformation has been shown to shift the ISCO inward under increasing
external magnetic field \cite{Siahaan:2026anp}.

On a parallel track, the study of black holes in theories beyond Einstein gravity
has been greatly stimulated by higher-dimensional and string-inspired frameworks.
Among the most natural extensions is Kaluza--Klein (KK) theory, in which one
extra spatial dimension is compactified to yield, in four dimensions, a metric
coupled to a $U(1)$ gauge field and a scalar dilaton. The resulting KK black
holes carry electric charge and dilaton hair absent in pure general relativity,
and their properties have been studied extensively
\cite{GarfinkleHorowitzStrominger1991,PonceDeLeonWesson1993,HorowitzTseytlin1994,
	MatsunoIshihara2008}. A closely related line of work by Ivashchuk and
collaborators has analysed dilatonic dyon and dyon-like black-hole solutions
in models with one or two Abelian gauge fields
\cite{Abishev:2015,Abishev:2017,Malybayev:2021}, together with their circular
geodesics and orbital stability
\cite{Boshkayev:2024dcg,Boshkayev:2025sdg}; the geodesic diagnostics computed
there for charged dilatonic backgrounds are natural points of comparison for
the present analysis. A particularly efficient technique for generating KK
solutions is the uplift--boost--reduction procedure \cite{Aliev:2008wv}: a
four-dimensional vacuum metric is lifted to five dimensions, a Lorentz boost is
performed along the extra dimension, and the resulting configuration is reduced
back to four dimensions. The boost parameter introduces an electric charge and a
nontrivial dilaton profile while leaving the causal structure of the seed metric
qualitatively intact. This method has been applied to the Schwarzschild and Kerr
seeds, and more recently to accelerating, Taub--NUT, and magnetized geometries
\cite{Siahaan:2024zhx,Siahaan:2024ilq,Siahaan:2024fbh}.

In this paper we apply the uplift--boost--reduction procedure to the
Zipoy--Voorhees seed and thereby construct the Kaluza--Klein
electric/dilaton counterpart of the Zipoy--Voorhees geometry, which we
call the Zipoy--Voorhees--dilaton (ZVD) spacetime. We emphasise that this is
the $a=\sqrt{3}$ Einstein--Maxwell--dilaton solution generated by the
five-dimensional boost technique; it is distinct from the charged ZV
spacetimes constructed in other Einstein--Maxwell or string-theory settings
\cite{Yunusov:2025chw}, which carry different scalar couplings and are not
related to ours by a field redefinition.
The new solution satisfies the equations of motion of four-dimensional
Kaluza--Klein theory and is parametrised by the seed length scale $M$, the
ZV deformation parameter $k$, and the boost parameter $\alpha$; the ADM
mass, electric charge, and dilaton charge are functions of these. For $k=1$
the spacetime is asymptotically flat and admits a regular black-hole
interpretation. For $k\neq1$ it inherits the naked curvature singularity of
the seed at $r=2M$ but remains asymptotically flat in the standard sense, as
shown explicitly in Section~\ref{sec.properties}. Setting $k=1$ reduces the
solution to the static KK black hole generated from the Schwarzschild seed,
while $\alpha=0$ recovers the uncharged Zipoy--Voorhees vacuum. We derive the
conserved charges, analyse the singular and horizon structure, and carry
out a detailed study of equatorial geodesics for neutral test particles
and photons, with particular emphasis on the photon ring and the
dependence of the innermost stable circular orbit on $k$ and $\alpha$.
Throughout the paper we restrict to the parameter ranges $k>0$, $\alpha\ge
0$, and to the exterior region $r>2M$; the explicit numerical analysis
covers $0.5\le k\le 1.5$ and $0\le\alpha\le 1.5$.

The remainder of the paper is organised as follows. In Section~\ref{sec.solution}
we introduce the Kaluza--Klein action and equations of motion, describe the
uplift--boost--reduction procedure, specify the Zipoy--Voorhees seed, and present
the full ZVD solution. Section~\ref{sec.properties} analyses the geometric and
asymptotic properties of the spacetime, including its limiting cases, singular
structure, and conserved charges; this section is placed before the geodesic
analysis so that the nature of the spacetime is established before orbits are
discussed. In Section~\ref{sec.geodesics} we restrict to the equatorial plane,
derive the effective potential for radial motion, obtain expressions for the
energy and angular momentum of circular orbits, and determine the photon ring and
ISCO; we then compute the gravitational redshift and the critical impact
parameter that sets the equatorial shadow scale. We
summarise our results and outline directions for future work in
Section~\ref{sec.conclusion}.
	
	\section{Zipoy--Voorhees--dilaton solution}\label{sec.solution}
	
	In the following, we consider two sets of fields. The first is the vacuum seed
	metric, denoted by $\{\tilde{g}_{\mu\nu}\}$, and the second is the corresponding
	four-dimensional Kaluza--Klein configuration, denoted by
	$\{g_{\mu\nu},A_\mu,\Phi\}$.
	
	The four-dimensional effective action of Kaluza--Klein theory is
	\cite{Aliev:2008wv}
	\[
	S=\int d^4x\,\sqrt{-g}\,
	\Bigl[
	R
	-2\,\partial_\mu\Phi\,\partial^\mu\Phi
	-e^{2\sqrt{3}\,\Phi}\,F_{\mu\nu}F^{\mu\nu}
	\Bigr]\,,
	\]
	where $R$ is the four-dimensional Ricci scalar, $\Phi$ the dilaton field, and
	\[
	F_{\mu\nu}=\partial_\mu A_\nu-\partial_\nu A_\mu
	\]
	the field strength of the $U(1)$ gauge potential $A_\mu$. Variation with respect
	to the metric, gauge field, and dilaton yields the equations of motion
	\be\label{eq.Rmn}
	R_{\mu\nu}
	=
	2\,\partial_\mu\Phi\,\partial_\nu\Phi
	+2\,e^{2\sqrt{3}\,\Phi}
	\!\left(
	F_{\mu\alpha}F_{\nu}{}^{\alpha}
	-\tfrac{1}{4}\,g_{\mu\nu}F_{\alpha\beta}F^{\alpha\beta}
	\right),
	\ee
	\be\label{eq.Am}
	\nabla_\mu\!\left(e^{2\sqrt{3}\,\Phi}F^{\mu\nu}\right)=0,
	\ee
	\be\label{eq.Phi}
	2\,\nabla^2\Phi-\sqrt{3}\,e^{2\sqrt{3}\,\Phi}F_{\alpha\beta}F^{\alpha\beta}=0.
	\ee
	
	A convenient way to construct solutions of this system is to start from a
	four-dimensional Ricci-flat seed metric $\tilde{g}_{\mu\nu}$ and uplift it to
	five dimensions as
	\be
	ds_5^2=\tilde{g}_{\mu\nu}\,dx^\mu dx^\nu+dz^2\,,
	\ee
	where $z$ is the extra coordinate. Since $z$ enters trivially, the
	five-dimensional metric is Ricci-flat whenever the seed satisfies
	$\tilde{R}_{\mu\nu}=0$. A nontrivial KK configuration is then generated by a
	Lorentz boost in the $(t,z)$-plane \cite{Aliev:2008wv},
	\[
	dt \to \cosh\alpha\,dt+\sinh\alpha\,dz,\qquad
	dz \to \cosh\alpha\,dz+\sinh\alpha\,dt,
	\]
	where $\alpha$ is the boost parameter. After the boost, the five-dimensional line
	element takes the standard KK form
	\be
	ds_5^2
	=
	H^{-1}\,g_{\mu\nu}\,dx^\mu dx^\nu
	+
	H^2\bigl(dz+2A_\mu\,dx^\mu\bigr)^2\,,
	\ee
	with
	\be\label{eq.H}
	H^2=\cosh^2\alpha+\tilde{g}_{tt}\sinh^2\alpha\,.
	\ee
	For a static seed ($\tilde{g}_{t\phi}=0$), the resulting four-dimensional fields
	are \cite{Siahaan:2024ilq}
	\be\label{eq.Agen}
	A_\mu\,dx^\mu
	=
	\frac{\sinh\alpha\cosh\alpha\,(1+\tilde{g}_{tt})}{2H^2}\,dt\,,
	\ee
	\be\label{eq.Phigen}
	\Phi=\frac{\sqrt{3}}{2}\ln H\,,
	\ee
	\be\label{eq.ds4gen}
	ds^2
	=
	\frac{\tilde{g}_{tt}}{H}\,dt^2
	+H\!\left(
	\tilde{g}_{rr}\,dr^2+\tilde{g}_{\theta\theta}\,d\theta^2
	+\tilde{g}_{\phi\phi}\,d\phi^2
	\right).
	\ee
	
	\subsection{The Zipoy--Voorhees seed}
	
	We take as our seed the Zipoy--Voorhees metric
	\cite{Zipoy:1966btu,Voorhees:1970ywo}, a static axisymmetric vacuum solution
	generalising Schwarzschild through a real deformation parameter $k$. Introducing
	\be\label{eq.metricFunctions}
	f(r)=1-\frac{2M}{r},
	\qquad
	h(r,\theta)=1-\frac{2M}{r}+\frac{M^2\sin^2\theta}{r^2},
	\ee
	the line element reads
	\be\label{eq.ZVseed}
	d\tilde{s}^2
	=
	-f^k\,dt^2
	+f^{k^2-k}h^{1-k^2}
	\!\left(
	\frac{dr^2}{f}+r^2\,d\theta^2
	\right)
	+r^2\sin^2\theta\,f^{1-k}\,d\phi^2.
	\ee
	The nonvanishing seed components are
	\be\label{eq.seedcomps}
	\tilde{g}_{tt}=-f^k,\quad
	\tilde{g}_{rr}=\frac{f^{k^2-k}h^{1-k^2}}{f},\quad
	\tilde{g}_{\theta\theta}=r^2 f^{k^2-k}h^{1-k^2},\quad
	\tilde{g}_{\phi\phi}=r^2\sin^2\theta\,f^{1-k}.
	\ee
	For $k=1$ the metric \eqref{eq.ZVseed} reduces to Schwarzschild, while $k\neq1$
	introduces a quadrupolar deformation breaking spherical symmetry.
	
	\subsection{The Zipoy--Voorhees--dilaton spacetime}
	
	Substituting \eqref{eq.seedcomps} into \eqref{eq.H},
	\be\label{eq.Hzv}
	H^2=\cosh^2\alpha-f^k\sinh^2\alpha
	=
	1+\bigl(1-f^k\bigr)\sinh^2\alpha.
	\ee
	The four-dimensional metric \eqref{eq.ds4gen} becomes
	\be\label{eq.ZVKKmetric}
	ds^2
	=
	-\frac{f^k}{H}\,dt^2
	+H\,\frac{f^{k^2-k}h^{1-k^2}}{f}\,dr^2
	+H\,r^2 f^{k^2-k}h^{1-k^2}\,d\theta^2
	+H\,r^2\sin^2\theta\,f^{1-k}\,d\phi^2,
	\ee
	the gauge potential \eqref{eq.Agen} reduces to a purely electric field,
	\be\label{eq.AZV}
	A_\mu\,dx^\mu
	=
	\frac{\sinh\alpha\cosh\alpha\,(1-f^k)}{2H^2}\,dt\,,
	\ee
	and the dilaton is
	\be\label{eq.PhiZV}
	\Phi=\frac{\sqrt{3}}{2}\ln H\,.
	\ee
	The fields \eqref{eq.ZVKKmetric}--\eqref{eq.PhiZV} define the Zipoy--Voorhees--dilaton (ZVD) spacetime. In
	the limit $\alpha\to0$ one recovers the vacuum Zipoy--Voorhees seed, whereas
	$k=1$ yields the static electrically charged KK black hole generated from the
	Schwarzschild seed.
	
	\section{Geometric properties and asymptotic charges}
	\label{sec.properties}
	
	\subsection{Limiting cases and singular structure}
	
	Two limiting cases are immediate. When $\alpha=0$, one has $H=1$, $A_\mu=0$,
	and the solution reduces to the vacuum Zipoy--Voorhees spacetime. When $k=1$,
	the seed becomes Schwarzschild and \eqref{eq.ZVKKmetric} reduces to the standard
	static electrically charged Kaluza--Klein black hole.
	
	For general $k\neq1$, however, the surface $r=2M$ inherited from the seed
	requires special care. From \eqref{eq.ZVKKmetric},
	\be\label{eq.gttgrr}
	g_{tt}=-\frac{f^k}{H},
	\qquad
	g_{rr}=H\,\frac{f^{k^2-k}h^{1-k^2}}{f},
	\ee
	so that
	\be\label{eq.product}
	g_{tt}g_{rr}=-f^{k^2-1}\,h^{1-k^2}.
	\ee
	For $k=1$ this product remains finite and nonzero at $f=0$, and one recovers the
	familiar Schwarzschild-type horizon structure. For $k\neq1$, by contrast,
	$g_{tt}g_{rr}$ either vanishes or diverges at $r=2M$ depending on the sign of
	$k^2-1$, indicating that the regular horizon structure is lost.
	
	This behaviour is consistent with the known properties of the vacuum
	Zipoy--Voorhees seed \cite{Herrera:1998rj}, for which curvature invariants diverge
	at $r=2M$ when $k\neq1$. Although the uplift--boost--reduction procedure modifies
	the four-dimensional metric through nontrivial warp factors, it does not
	regularise this singular surface. Hence for $k\neq1$ the ZVD spacetime inherits
	a naked curvature singularity at $r=2M$ rather than a regular horizon.
	
	The regularity of the symmetry axis may also be examined directly. The relevant
	criterion is the elementary-flatness ratio,
	\[
	F(r,\theta)\equiv\frac{g_{\phi\phi}}{g_{\theta\theta}\sin^2\theta}
	\qquad\text{as }\theta\to0,\pi.
	\]
	In the ZVD geometry, the Kaluza--Klein transformation modifies the angular
	sector only through the common conformal factor $H$, which cancels in this
	ratio. Substituting the explicit components in \eqref{eq.ZVKKmetric} and
	using \eqref{eq.metricFunctions}, the ratio reduces to
	\begin{equation}\label{eq.flatratio}
	F(r,\theta)=\frac{f^{1-k}}{f^{k^2-k}h^{1-k^2}}=\left(\frac{f}{h}\right)^{1-k^2}.
	\end{equation}
	On the symmetry axis $\theta\to 0,\pi$ at any $r>2M$ one has
	$h(r,\theta)\to f(r)$ exactly (since the term $M^2\sin^2\theta/r^2$ in
	\eqref{eq.metricFunctions} vanishes), and consequently $F\to 1$. The axis
	is therefore locally regular for every $k$ throughout the exterior $r>2M$,
	and the ZVD spacetime carries no conical defect on the half-axis away from
	the singular surface. This is consistent with the standard treatment of the
	vacuum Zipoy--Voorhees ($\gamma$-) metric in Schwarzschild-like coordinates;
	any conical-defect features that arise in Weyl $(\rho,z)$ coordinates are
	concentrated on the singular rod at $r=2M$ and do not extend to spatial
	infinity \cite{Herrera:1998rj,Kodama:2003ch}.
	
	The asymptotic structure is similarly clean. Expanding the metric components
	at large $r$ at fixed $\theta$,
	\[
	g_{tt}\to -1,\qquad
	g_{rr}\to 1,\qquad
	\frac{g_{\theta\theta}}{r^2}\to 1,\qquad
	\frac{g_{\phi\phi}}{r^2\sin^2\theta}\to 1,
	\]
	with subleading terms of order $1/r$. The metric on a constant-$r$,
	constant-$t$ surface tends to the round 2-sphere of area $4\pi r^2$, and
	the spacetime is asymptotically flat in the standard ADM sense for every
	$k$. The ADM charges derived in the next subsection are therefore the
	conventional ones, with the standard round-sphere integration measure.
	
	Accordingly, for $k\neq 1$ the ZVD solution describes a static,
	asymptotically flat, charged dilatonic axisymmetric spacetime with a naked
	curvature singularity at $r=2M$ as its only nontrivial geometric feature
	apart from the quadrupole deformation. The black-hole interpretation
	applies only in the spherical limit $k=1$.
	
	\subsection{Asymptotic charges}
	
	The asymptotic conserved charges follow from the large-$r$ expansion of the
	metric, gauge field, and dilaton. From $f(r)=1-2M/r$,
	\be\label{eq.fexpand}
	f^k=1-\frac{2kM}{r}+\mathcal{O}(r^{-2}).
	\ee
	Using \eqref{eq.Hzv}, this implies
	\be\label{eq.Hexpand}
	H^2=1+\frac{2kM\sinh^2\alpha}{r}+\mathcal{O}(r^{-2}),
	\qquad
	H=1+\frac{kM\sinh^2\alpha}{r}+\mathcal{O}(r^{-2}).
	\ee
	Therefore
	\be\label{eq.gttexpand}
	g_{tt}
	=-\frac{f^k}{H}
	=-1+\frac{kM(1+\cosh^2\alpha)}{r}+\mathcal{O}(r^{-2}).
	\ee
	Comparing with the asymptotically flat form
	$g_{tt}=-1+2\mathcal{M}/r+\mathcal{O}(r^{-2})$,
	we identify the ADM mass
	\be\label{eq.ADMmass}
	\mathcal{M}=\frac{kM}{2}\bigl(1+\cosh^2\alpha\bigr).
	\ee
	Since the seed is static and the boost does not generate a four-dimensional
	rotation, the angular momentum vanishes:
	\be\label{eq.Jcharge}
	J=0.
	\ee
	The large-$r$ expansion of the gauge potential \eqref{eq.AZV} gives
	\be\label{eq.Atexpand}
	A_t=\frac{kM\sinh\alpha\cosh\alpha}{r}+\mathcal{O}(r^{-2}),
	\ee
	from which the electric charge is
	\be\label{eq.charge}
	Q=kM\sinh\alpha\cosh\alpha=\frac{kM}{2}\sinh 2\alpha.
	\ee
	Finally, the dilaton expands as
	\be\label{eq.Phiexpand}
	\Phi=\frac{\sqrt{3}\,kM\sinh^2\alpha}{2r}+\mathcal{O}(r^{-2}),
	\ee
	giving the dilaton charge
	\be\label{eq.dilatoncharge}
	\Sigma=\frac{\sqrt{3}}{2}\,kM\sinh^2\alpha.
	\ee
	Collecting these results, the three asymptotic charges are
	\[
	\mathcal{M}=\frac{kM}{2}(1+\cosh^2\alpha),\qquad
	Q=\frac{kM}{2}\sinh 2\alpha,\qquad
	\Sigma=\frac{\sqrt{3}}{2}\,kM\sinh^2\alpha.
	\]
	All three are determined by the two parameters $M$ and $\alpha$ once $k$ is
	fixed. The limits $\alpha=0$ and $k=1$ reproduce, respectively, the vacuum
	Zipoy--Voorhees seed and the standard static charged Kaluza--Klein solution.
	
	Although $\mathcal{M}$, $Q$, and $\Sigma$ appear as three separate charges, the
	dilaton charge is not independent. A direct computation shows that they satisfy
	the constraint
	\be\label{eq.dilatonconstraint}
	\Sigma^2+\sqrt{3}\,\mathcal{M}\,\Sigma=\frac{3}{2}\,Q^2,
	\ee
	for all values of $k$ and $\alpha$. The reason for this $k$-independence is
	transparent: comparing \eqref{eq.ADMmass}, \eqref{eq.charge}, and
	\eqref{eq.dilatoncharge} with the corresponding charges of the
	$k=1$ Schwarzschild--Kaluza--Klein solution, one sees that the ZVD charges are
	simply $k$ times the latter,
	\[
	\bigl(\mathcal{M},\,Q,\,\Sigma\bigr)\bigl|_{\rm ZVD}
	=k\,\bigl(\mathcal{M},\,Q,\,\Sigma\bigr)\bigl|_{k=1}.
	\]
	In other words the quadrupole deformation rescales the effective mass parameter
	from $M$ to $kM$ but leaves all three charges in the same ratios. Since
	\eqref{eq.dilatonconstraint} is a homogeneous quadratic identity in
	$(\mathcal{M},Q,\Sigma)$, it lifts automatically from the $k=1$ relation
	\cite{Aliev:2008wv} to the entire ZVD family, and the dilaton charge is
	determined by the ADM mass and the electric charge regardless of the quadrupole
	deformation.
	
	It is worth pausing on the physical meaning of the boost parameter $\alpha$,
	which controls all of the charges above. By construction $\alpha$ is the
	\emph{rapidity} of the Lorentz boost performed in the $(t,z)$-plane of the
	five-dimensional uplift [the transformation preceding \eqref{eq.H}]. This is
	precisely why it enters every expression only through the hyperbolic functions
	$\cosh\alpha$ and $\sinh\alpha$ and behaves as a (hyperbolic) boost angle
	rather than as a length or a charge: a rapidity is dimensionless and additive,
	and $\alpha\to0$ is the unboosted seed. In the five-dimensional picture the
	static seed carries momentum purely along $\partial_t$; boosting by rapidity
	$\alpha$ rotates a component of this momentum into the compact direction
	$\partial_z$, and upon Kaluza--Klein reduction that internal momentum is
	exactly the four-dimensional electric charge \eqref{eq.charge}, while the
	accompanying change in $\tilde g_{tt}$ sources the dilaton \eqref{eq.PhiZV}.
	Because the boost is a solution-generating transformation, distinct values of
	$\alpha$ label genuinely inequivalent four-dimensional spacetimes rather than
	the same geometry in different coordinates, so it is natural that the
	asymptotic charges depend on it. Quantitatively, $\alpha$ acts as the
	charge-to-mass dial of the configuration: from \eqref{eq.ADMmass} and
	\eqref{eq.charge},
	\be\label{eq.QoverM}
	\frac{Q}{\mathcal{M}}=\frac{\sinh 2\alpha}{1+\cosh^2\alpha},
	\ee
	which is independent of $k$, vanishes at $\alpha=0$ (the uncharged ZV seed),
	increases monotonically with $\alpha$, and approaches the Kaluza--Klein bound
	$Q/\mathcal{M}\to2$ as $\alpha\to\infty$. The physically admissible range of
	electric charge is thus swept out by $\alpha\in[0,\infty)$, the surveyed
	window $0\le\alpha\le1.5$ corresponding to $0\le Q/\mathcal{M}\lesssim1.5$.
	
	\section{Equatorial geodesics and circular motion}
	\label{sec.geodesics}
	
	The geodesic structure of Zipoy--Voorhees-type geometries is considerably richer
	than in the Schwarzschild case \cite{Herrera:1998rj}. We restrict here to motion
	on the symmetry plane $\theta=\pi/2$, which is sufficient for the analysis of
	circular orbits and their stability. The analysis below concerns
	\emph{neutral} test particles (with $\mu_0=1$) and photons (with $\mu_0=0$),
	which follow geodesics of \eqref{eq.ZVKKmetric}; charged test particles
	would couple to the gauge potential \eqref{eq.AZV} and obey Lorentz-force
	equations of motion, modifying the conserved quantities by additional
	$qA_\mu$ contributions, and we do not consider that case here.
	Throughout this section we set $M=1$ (so that radii are quoted in units of
	$M$ and frequencies in units of $1/M$); the dimensionful Schwarzschild
	benchmark $\Omega_{\rm ISCO}=1/(6\sqrt{6}\,M)$ is restored in the obvious
	way. The relation to the ADM mass
	$\mathcal{M}=\tfrac{kM}{2}(1+\cosh^2\alpha)$ is discussed at the end of the
	section.
	
	As established in Section~\ref{sec.properties}, for $k\neq1$ the surface $r=2M$
	is a naked curvature singularity rather than a regular horizon. In the
	test-body approximation adopted here the geodesic equations make sense
	throughout $r>2M$, and circular orbits in particular can be studied as
	solutions of an ordinary algebraic system. The analysis is most straightforward
	when the relevant orbits lie at radii where curvature invariants remain
	bounded, i.e.\ not too close to $r=2M$. Where we discuss circular orbits whose
	radii approach the singular surface --- which happens for the photon ring near
	$k\simeq\tfrac{1}{2}$, see Section~\ref{sec.null} --- the corresponding test
	particles probe a region in which the Kretschmann scalar of the underlying
	geometry diverges \cite{Herrera:1998rj}, and the formal solutions of the
	geodesic equations there should be interpreted with appropriate care: they
	describe extremal mathematical orbits in a singular background rather than
	physically realisable trajectories of an extended body. With this caveat the
	geodesic structure is well-defined and its qualitative dependence on $k$ and
	$\alpha$ is the subject of the present section.
	
	The metric \eqref{eq.ZVKKmetric} is static and axisymmetric, admitting the two
	Killing vectors $\partial_t$ and $\partial_\phi$. These give rise to two conserved
	quantities along any geodesic: the specific energy $E$ and the specific azimuthal
	angular momentum $L$,
	\begin{equation}\label{eq.Edef}
	E=-g_{tt}\dot{t},\qquad L=g_{\phi\phi}\dot{\phi},
	\end{equation}
	where an overdot denotes differentiation with respect to an affine parameter
	$\lambda$. The tangent vector satisfies
	\begin{equation}\label{eq.norm}
	g_{\mu\nu}\dot{x}^\mu\dot{x}^\nu=-\mu_0^2,
	\end{equation}
	with $\mu_0=1$ for timelike and $\mu_0=0$ for null geodesics.
	
	\subsection{Equatorial motion}
	
	Owing to the reflection symmetry $\theta\to\pi-\theta$, the equatorial plane is
	an invariant submanifold: a geodesic starting at $\theta=\pi/2$ with
	$\dot{\theta}=0$ remains there. On $\theta=\pi/2$,
	\begin{equation}\label{eq.heq}
	h_e(r)\equiv h(r,\pi/2)
	=1-\frac{2M}{r}+\frac{M^2}{r^2}
	=\left(1-\frac{M}{r}\right)^2.
	\end{equation}
	The induced equatorial line element is
	\begin{equation}\label{eq.eqmetric}
	ds^2_{\rm eq}=-A(r)\,dt^2+B(r)\,dr^2+C(r)\,d\phi^2,
	\end{equation}
	where
	\begin{equation}\label{eq.ABCdef}
	A(r)=\frac{f^k}{H},\quad
	B(r)=H\,f^{k^2-k-1}\,h_e^{1-k^2},\quad
	C(r)=H\,r^2 f^{1-k},
	\end{equation}
	and $H^2=\cosh^2\alpha-f^k\sinh^2\alpha$. Hence
	\begin{equation}\label{eq.Eeq}
	E=A(r)\,\dot{t},\qquad L=C(r)\,\dot{\phi}.
	\end{equation}
	Substituting into \eqref{eq.norm},
	\begin{equation}\label{eq.radialeq}
	B(r)\,\dot{r}^2=\frac{E^2}{A(r)}-\frac{L^2}{C(r)}-\mu_0^2
	\equiv\mathcal{R}(r;E,L,\mu_0),
	\end{equation}
	or equivalently
	\begin{equation}\label{eq.radialeq2}
	A(r)B(r)\,\dot{r}^2+V_{\rm eff}(r;L,\mu_0)=E^2,
	\end{equation}
	with the effective potential
	\begin{equation}\label{eq.Veff}
	V_{\rm eff}(r;L,\mu_0)=A(r)\!\left(\mu_0^2+\frac{L^2}{C(r)}\right).
	\end{equation}
	
	For numerical illustration it is convenient to set $M=1$ and introduce the
	dimensionless variables $x=r/M$ and $\ell=L/(\mu_0 M)$, writing
	\begin{equation}\label{eq.Vtilde}
	\tilde{V}_{\rm eff}(x)\equiv\frac{V_{\rm eff}}{\mu_0^2}
	=\frac{f(x)^k}{H(x)}
	+\frac{\ell^2 f(x)^{2k-1}}{H(x)^2\,x^2},\quad
	H^2(x)=\cosh^2\alpha-f(x)^k\sinh^2\alpha,
	\end{equation}
	with $f(x)=1-2/x$. One has $\tilde{V}_{\rm eff}\to1$ as $x\to\infty$, so bound
	timelike motion is possible whenever the potential develops a local minimum below
	unity. As a benchmark, in the Schwarzschild limit $(k,\alpha)=(1,0)$ with
	$\ell=4$, there is an unstable circular orbit at $x=4$ with
	$\tilde{V}_{\rm eff}=1$ and a stable one at $x=12$ with
	$\tilde{V}_{\rm eff}=25/27\simeq0.9259$.
	
\begin{figure*}[t]
		\centering
		\begin{subfigure}[t]{0.32\textwidth}
			\centering
			\includegraphics[width=\textwidth]{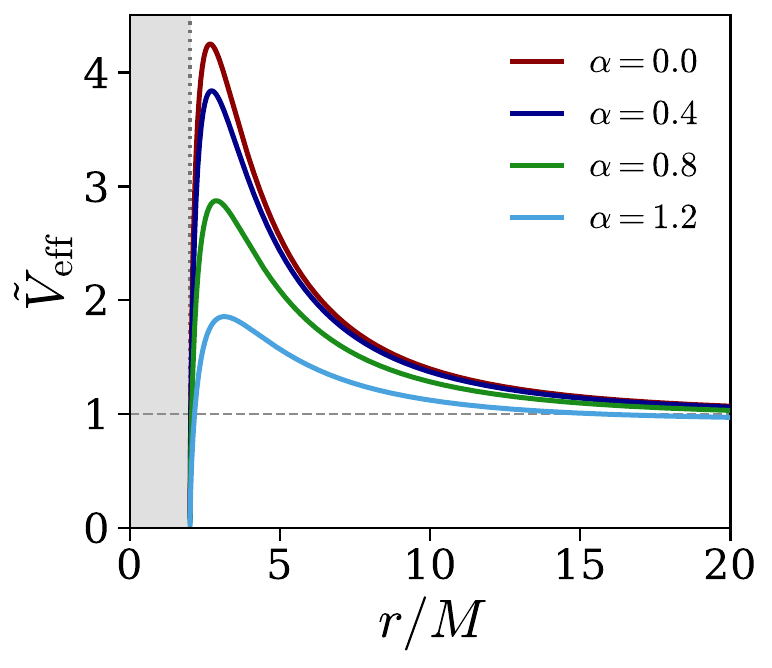}
			\caption{$k=0.8$}
			\label{fig:veff_k08}
		\end{subfigure}
		\hfill
		\begin{subfigure}[t]{0.32\textwidth}
			\centering
			\includegraphics[width=\textwidth]{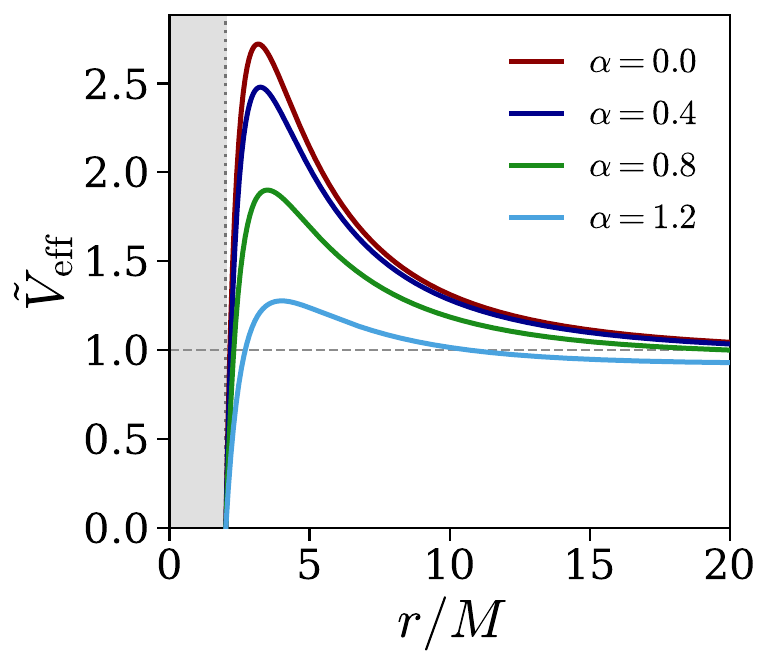}
			\caption{$k=1.0$}
			\label{fig:veff_k10}
		\end{subfigure}
		\hfill
		\begin{subfigure}[t]{0.32\textwidth}
			\centering
			\includegraphics[width=\textwidth]{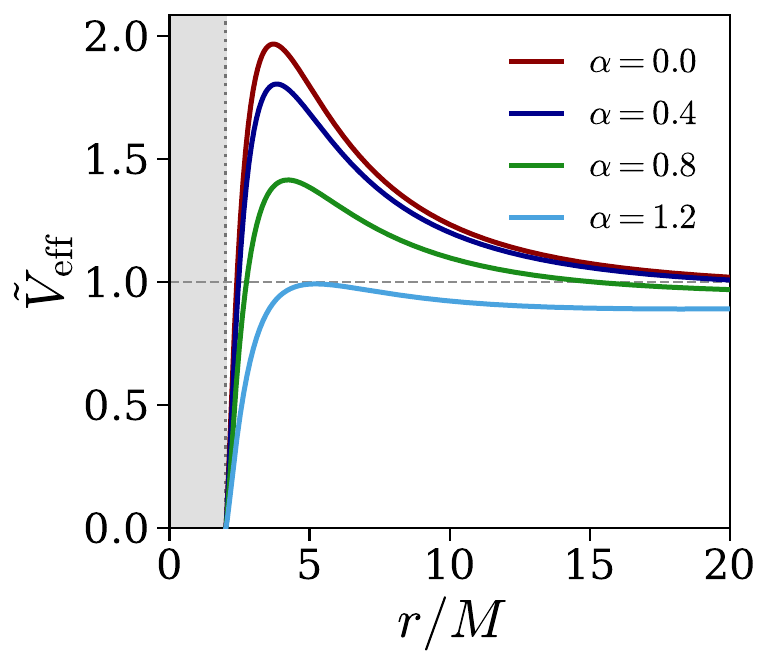}
			\caption{$k=1.2$}
			\label{fig:veff_k12}
		\end{subfigure}
		\caption{Dimensionless effective potential
			$\tilde{V}_{\rm eff}=V_{\rm eff}/\mu_0^2$ for equatorial timelike motion
			in the ZVD spacetime, shown for $k=0.8$, $1.0$, and $1.2$ with $M=\mu_0=1$
			and fixed specific angular momentum $\ell=L/(\mu_0 M)=8$. In each panel,
			the four curves correspond to $\alpha=0.0$ (dark red), $0.4$ (dark blue),
			$0.8$ (green), and $1.2$ (light blue), in order of decreasing peak
			height. The horizontal axis starts at $r=0$: the shaded strip
			$r\le 2M$ lies outside the exterior domain of the solution, and the
			vertical dotted line marks the surface $r=2M$, which is a regular
			horizon only for $k=1$ and a naked curvature singularity otherwise.
			The horizontal dashed line marks $\tilde{V}_{\rm eff}=1$, which
			separates bound from unbound radial motion. Note that
			$\tilde{V}_{\rm eff}\to0$ as $r\to2M^{+}$ for all $k>1/2$, so the
			barrier is of finite height and the vertical scale is chosen
			separately in each panel so that its full height is visible.
			Where present, the inner local maximum marks an unstable
			circular orbit and the outer local minimum marks a stable one; for
			$k=0.8$ at small $\alpha$ the outer minimum lies beyond the plotted
			range and only the centrifugal barrier is visible. Increasing $\alpha$
			at fixed $k$ lowers the barrier height and draws the potential well
			inward; increasing $k$ at fixed $\alpha$ shifts the inner maximum
			outward and deepens the well, with the combined effect most pronounced
			for large $\alpha$ and large $k$.}
		\label{fig:veff_sidebyside}
	\end{figure*}

	Representative extrema of the potential are listed in Table~\ref{tab:Veff_num}.
	Both $k$ and $\alpha$ substantially affect the location of the inner barrier and
	the depth of the outer well. Figure~\ref{fig:veff_sidebyside} presents
	$\tilde{V}_{\rm eff}$ for $k=0.8$, $1.0$, and $1.2$, with $M=\mu_0=1$,
	$\ell=8$, and $\alpha=0.0,0.4,0.8,1.2$.
	
	\begin{table}[t]
		\centering
		\begin{tabular}{cccccc}
			\hline
			$(k,\alpha,\ell)$ & $x_{\rm max}$ & $\tilde{V}_{\rm eff}(x_{\rm max})$
			& $x_{\rm min}$ & $\tilde{V}_{\rm eff}(x_{\rm min})$ & interpretation \\
			\hline
			$(1,0,4)$     & $4.000$ & $1.0000$ & $12.000$ & $0.9259$
			& Schwarzschild benchmark \\
			$(1,0.8,8)$   & $3.487$ & $1.8993$ & $42.046$ & $0.9683$
			& shallower outer well \\
			$(1.2,0.8,8)$ & $4.218$ & $1.4147$ & $33.210$ & $0.9533$
			& well moves inward \\
			$(1.3,0.8,8)$ & $4.651$ & $1.2599$ & $29.608$ & $0.9444$
			& deeper, more localized minimum \\
			$(1.2,1.2,8)$ & $5.208$ & $0.9922$ & $18.615$ & $0.8901$
			& strong deepening of the well \\
			\hline
		\end{tabular}
		\caption{Representative extrema of the dimensionless effective potential
			$\tilde{V}_{\rm eff}=V_{\rm eff}/\mu_0^2$ for equatorial timelike motion
			($M=\mu_0=1$, $x=r/M$, $\ell=L/(\mu_0 M)$). Here $x_{\rm max}$ denotes
			the inner local maximum (unstable circular orbit) and $x_{\rm min}$ the
			outer local minimum (stable circular orbit), where both extrema exist within
			the plotted range.}
		\label{tab:Veff_num}
	\end{table}
	
	\subsection{Circular timelike orbits}
	
	A circular timelike orbit at $r=r_c$ is characterised by
	\begin{equation}\label{eq.circconds}
	\mathcal{R}(r_c)=0,\qquad\mathcal{R}'(r_c)=0,
	\end{equation}
	or equivalently by $\dot{r}|_{r_c}=0$ and $dV_{\rm eff}/dr|_{r_c}=0$.
	For the static equatorial metric \eqref{eq.eqmetric} these conditions yield
	\begin{equation}\label{eq.Omega}
	\Omega^2\equiv\left(\frac{d\phi}{dt}\right)^2=\frac{A'(r_c)}{C'(r_c)},
	\end{equation}
	\begin{equation}\label{eq.Ec}
	E^2=\frac{A(r_c)^2\,C'(r_c)}{A(r_c)\,C'(r_c)-C(r_c)\,A'(r_c)},
	\end{equation}
	\begin{equation}\label{eq.Lc}
	L^2=\frac{C(r_c)^2\,A'(r_c)}{A(r_c)\,C'(r_c)-C(r_c)\,A'(r_c)}.
	\end{equation}
	It is useful to define
	\begin{equation}\label{eq.Ddef}
	D(r)\equiv A(r)\,C'(r)-C(r)\,A'(r),
	\end{equation}
	so that the existence of a timelike circular orbit requires $D(r_c)>0$, together
	with $E^2>0$ and $L^2>0$.
	
	For the numerical analysis we set $M=1$ and write
	\begin{equation}
	A(r)=\frac{f(r)^k}{H(r)},\qquad C(r)=H(r)\,r^2 f(r)^{1-k},
	\end{equation}
	which satisfy the exact identity
	\begin{equation}
	A(r)\,C(r)=r(r-2).
	\end{equation}
	Introducing
	\begin{equation}
	\Upsilon(r)\equiv k\!\left(2+\frac{f(r)^k\sinh^2\alpha}{H(r)^2}\right),
	\end{equation}
	one finds
	\begin{equation}
	D(r)=2\bigl[r-1-\Upsilon(r)\bigr].
	\end{equation}
	The behaviour of $D(r)$ near the singular surface is easily characterised
	analytically. As $r\to 2M^+$ one has $f\to 0^+$, hence $f^k\to 0$ and
	$H^2\to\cosh^2\alpha$, so $\Upsilon(r)\to 2k$ and
	\begin{equation}\label{eq.Datsing}
	D(r)\;\longrightarrow\;2\bigl[r-1-2k\bigr]\Big|_{r=2}
	\;=\;2\bigl(1-2k\bigr)\qquad\text{as}\qquad r\to 2M^+.
	\end{equation}
	The sign of the limit changes at $k=1/2$, and since $D(r)\to+\infty$ as
	$r\to\infty$, $D(r)$ has at least one zero in $(2M,\infty)$ whenever
	$k>1/2$. The number of exterior zeros is not inferred from a global
	monotonicity argument; indeed, an explicit computation gives
	\begin{equation}\label{eq.dUpsilon}
	\frac{d\Upsilon}{dr}=\frac{2k^2\sinh^2\alpha\cosh^2\alpha}
	{r^2 H^4}\,f^{k-1},
	\end{equation}
	which diverges as $r\to 2M^+$ for $0<k<1$ and $\alpha\neq 0$, so $D'(r)$
	cannot be positive throughout $r>2M$ in that regime. Rather than invoke
	monotonicity, we determine the number of exterior roots numerically over
	the surveyed domain $0.5\le k\le 1.5$, $0\le\alpha\le 1.5$ on a grid of
	resolution $\Delta k=\Delta\alpha=0.025$. In every sampled point with
	$k>1/2$ a single exterior root is found (located by sign-change scan
	followed by bracketed root refinement to tolerance $10^{-6}$), and in
	every sampled point with $k<1/2$ no exterior root is found. Where
	$D'(r)$ does become negative near $r=2M^+$, the resulting non-monotonic
	dip is shallow and remains below zero, so $D$ still crosses zero only
	once on its way to $+\infty$.
	
	Combining \eqref{eq.Datsing} with the numerical root counting, the
	photon-ring structure organises into three regimes:
	\begin{itemize}
		\item For $k>1/2$, $D(2M^+)<0$ while $D(\infty)>0$, and a unique
		exterior root $r_{\rm ph}>2M$ is found in the parameter survey;
		it determines the photon-ring radius and simultaneously serves as
		the inner edge of the timelike circular family (see
		Section~\ref{sec.null} below). The threshold $k>1/2$ generalises a
		known feature of the vacuum Zipoy--Voorhees ($\gamma$-) metric
		\cite{Kodama:2003ch}, where the exterior photon orbit at
		$r_{\rm ph}=(1+2\gamma)M$ ceases to exist below $\gamma=1/2$. Our
		analysis shows that the boost parameter $\alpha$ shifts
		$r_{\rm ph}$ outward but does not alter the threshold, and
		provides the explicit elementary expression \eqref{eq.Datsing} for
		the location of the boundary in the electrically charged case. The
		window $1/2<k<1$ deserves emphasis: an exterior photon ring exists
		there even though the spacetime carries a naked singularity rather
		than a regular horizon.
		\item For $k=1/2$, $D(2M^+)=0$ at leading order; the subleading
		correction to $D$ near $r=2M^+$ is governed by the boost-induced
		term $-2k\tanh^2\alpha\,f^k$, which is negative for $\alpha\neq 0$
		and vanishes for $\alpha=0$. The exterior root is therefore
		generated by the boost: at $\alpha=0$ it sits exactly at the
		singular surface, while for $\alpha>0$ it lies just outside.
		The representative case $(k,\alpha)=(0.5,0.5)$ has
		$r_{\rm ph}\simeq 2.006M$, very close to the singular surface; for
		larger $\alpha$ the root moves further outward (reaching, e.g.,
		$r_{\rm ph}\simeq 2.12M$ at $\alpha=1.5$) but still remains in the
		near-singularity region in the surveyed range. Because the
		associated photon orbit grazes a region of unbounded curvature, we
		distinguish it from the physically cleaner $k>1/2$ photon ring.
		\item For $k<1/2$, $D(2M^+)>0$, and the numerical scan finds no
		exterior root: $D(r)>0$ throughout $r>2M$ in every sampled point.
		No exterior photon ring forms, and the lower boundary of the
		timelike circular-orbit domain coincides with the naked
		singularity at $r=2M$.
	\end{itemize}
	When an exterior photon ring exists we use $r_{\rm ph}$ for its
	radius; in the third case we set $r_{\rm ph}=2M$ as a formal lower bound.
	
	The marginally bound orbit $r_{\rm mb}$ is determined by $E(r)^2=1$, and the
	ISCO by
	\begin{equation}\label{eq.ISCOcond}
	\frac{dL^2}{dr}=0,\quad\text{equivalent to}\quad V_{\rm eff}''(r)=0.
	\end{equation}
Our numerical survey for $0.5\le k\le1.5$ and $0\le\alpha\le1.5$ reveals a single
	connected family of equatorial timelike circular orbits with the ordering
	\begin{equation}
	r_{\rm ph}<r_{\rm mb}<r_{\rm ISCO}.
	\end{equation}
	The segment $r_{\rm ph}<r<r_{\rm ISCO}$ is unstable; $r>r_{\rm ISCO}$ is stable.
	The Schwarzschild limit $(k,\alpha)=(1,0)$ gives
	\begin{equation}
	r_{\rm ph}=3,\qquad r_{\rm mb}=4,\qquad r_{\rm ISCO}=6.
	\end{equation}
	Representative values are listed in Table~\ref{tab:circular_timelike}. At fixed
	$k$, increasing $\alpha$ shifts all characteristic radii outward; at fixed
	$\alpha=0.5$, increasing $k$ pushes them outward more strongly, with
	$r_{\rm ISCO}$ moving from $3.133$ at $k=0.5$ to $9.054$ at $k=1.5$.
	
	\begin{table}[t]
		\centering
		\caption{Representative equatorial timelike circular orbits for the ZVD
			spacetime (units $M=1$). For $k>1/2$, $r_{\rm ph}$ is the unique
			genuine exterior zero of $D(r)$ and coincides with the photon-ring
			radius. For $k=1/2$ and $\alpha>0$, a formal near-singularity root
			exists just outside $r=2M$; because the corresponding photon orbit
			probes the curvature singularity we distinguish it from the
			physically cleaner $k>1/2$ case (see Section~\ref{sec.null}). For
			$k<1/2$ no exterior root exists and $r_{\rm ph}$ then denotes the
			lower boundary of the circular-orbit domain at the naked singularity.
			$r_{\rm mb}$ is defined by $E^2=1$ and $r_{\rm ISCO}$ by
			$dL^2/dr=0$.}
		\label{tab:circular_timelike}
		\begin{tabular}{cccccccc}
			\hline
			$k$ & $\alpha$ & $r_{\rm ph}$ & $r_{\rm mb}$ & $r_{\rm ISCO}$
			& $E_{\rm ISCO}$ & $L_{\rm ISCO}$ & $\Omega_{\rm ISCO}$ \\
			\hline
			0.5 & 0.5 & 2.006 & 2.307 & 3.133 & 0.928 & 1.775 & 0.176 \\
			0.8 & 0.5 & 2.660 & 3.438 & 5.074 & 0.938 & 3.001 & 0.084 \\
			1.0 & 0.0 & 3.000 & 4.000 & 6.000 & 0.943 & 3.464 & 0.068 \\
			1.0 & 0.5 & 3.081 & 4.116 & 6.236 & 0.939 & 3.787 & 0.065 \\
			1.0 & 1.0 & 3.295 & 4.361 & 6.779 & 0.924 & 4.875 & 0.058 \\
			1.2 & 0.5 & 3.500 & 4.781 & 7.372 & 0.940 & 4.568 & 0.053 \\
			1.5 & 0.5 & 4.129 & 5.767 & 9.054 & 0.941 & 5.732 & 0.042 \\
			\hline
		\end{tabular}
	\end{table}
	
	\subsection{Null circular orbits}
	\label{sec.null}
	
	For null geodesics ($\mu_0=0$), the impact parameter $b\equiv L/E$ is conserved.
	A circular null orbit satisfies
	\begin{equation}\label{eq.nullcirc1}
	\mathcal{R}(r_{\rm ph})=0,\qquad\mathcal{R}'(r_{\rm ph})=0,
	\end{equation}
	which is equivalent to
	\begin{equation}\label{eq.nullcirc}
	\frac{d}{dr}\!\left(\frac{C(r)}{A(r)}\right)\bigg|_{r=r_{\rm ph}}=0.
	\end{equation}
	The critical impact parameter at such a radius is
	\begin{equation}\label{eq.bph}
	b_{\rm ph}^2=\frac{C(r_{\rm ph})}{A(r_{\rm ph})}.
	\end{equation}
	Using the identity
	\begin{equation}
	\frac{d}{dr}\!\left(\frac{C}{A}\right)=\frac{D(r)}{A(r)^2},
	\end{equation}
where $D(r)$ is defined in (\ref{eq.Ddef}),	the null circular-orbit condition reduces simply to
	\begin{equation}\label{eq.nullDzero}
	D(r_{\rm ph})=0,
	\end{equation}
	which coincides with the inner-edge condition of the timelike circular family.
	In terms of $\Upsilon(r)$ defined above, the photon-ring condition reads
	\begin{equation}
	r_{\rm ph}=1+k\!\left(2+\frac{f(r_{\rm ph})^k\sinh^2\alpha}{H(r_{\rm ph})^2}\right).
	\end{equation}
	The orbital frequency at the photon ring is
	\begin{equation}
	\Omega_{\rm ph}=\sqrt{\frac{A(r_{\rm ph})}{C(r_{\rm ph})}}=\frac{1}{b_{\rm ph}},
	\end{equation}
	generalising the Schwarzschild values
	\begin{equation}
	r_{\rm ph}=3,\qquad b_{\rm ph}=3\sqrt{3},\qquad\Omega_{\rm ph}=\frac{1}{3\sqrt{3}},
	\end{equation}
	recovered at $(k,\alpha)=(1,0)$.
	
	The threshold for the existence of a genuine exterior photon ring follows
	directly from \eqref{eq.Datsing}. The exterior root of $D(r)=0$ exists for
	every $k>1/2$, in particular throughout the region $1/2<k<1$ where the
	spacetime contains a naked singularity rather than a regular horizon: an
	exterior photon ring is therefore not a black-hole-only feature in the ZVD
	family. For $k=1/2$ at $\alpha\neq 0$, the leading limit \eqref{eq.Datsing}
	vanishes and the subleading term $-2k\tanh^2\alpha\,f^k$ pulls $D$ slightly
	negative just above $r=2M$, so the root exists but lies in the
	near-singularity region: at $\alpha=0.5$ it is at
	$r_{\rm ph}\simeq 2.006M$, moving outward with $\alpha$ to
	$r_{\rm ph}\simeq 2.12M$ at $\alpha=1.5$ but staying close to the
	singular surface throughout the surveyed range. The associated photon
	orbit therefore probes a region of unbounded curvature and is of
	limited physical relevance. For $k<1/2$, the exterior root disappears
	altogether: the parameter survey finds $D(r)>0$ throughout $r>2M$, no
	exterior photon ring forms, and the lower boundary of the timelike
	circular-orbit family coincides with the singularity. Whenever an
	exterior photon ring does exist, we find
	\begin{equation}
	\frac{d^2}{dr^2}\!\left(\frac{C}{A}\right)\bigg|_{r=r_{\rm ph}}>0,
	\end{equation}
	equivalently a maximum of the null effective potential $A/C$, confirming that
	the null circular orbit is unstable. Representative numerical values are given
	in Table~\ref{tab:null_circular}. Both $k$ and $\alpha$ increase the
	photon-ring radius and the critical impact parameter, enlarging the photon
	capture region and reducing $\Omega_{\rm ph}$.
	
	\begin{table}[t]
		\centering
		\caption{Representative null circular orbits for the ZVD spacetime (units
			$M=1$). For $k>1/2$ a genuine exterior root of $D(r)=0$ exists and
			defines the photon ring; the corresponding orbit is unstable in every
			case. For $k=0.5$ at $\alpha\neq0$ the root sits very close to the
			singular surface $r=2M$ (here at $r_{\rm ph}\simeq 2.006$) and the
			associated photon orbit is therefore of limited physical relevance,
			although it remains formally an unstable null circular orbit. For
			$k<1/2$ no exterior root exists and a dash is shown.}
		\label{tab:null_circular}
		\begin{tabular}{cccccc}
			\hline
			$k$ & $\alpha$ & $r_{\rm ph}$ & $b_{\rm ph}$ & $\Omega_{\rm ph}$
			& Stability \\
			\hline
			0.3 & 0.5 & --    & --    & --    & no exterior ring \\
			0.5 & 0.5 & 2.006 & 2.249 & 0.445 & unstable (near-singularity) \\
			0.6 & 0.5 & 2.235 & 3.068 & 0.326 & unstable \\
			0.8 & 0.5 & 2.660 & 4.394 & 0.228 & unstable \\
			1.0 & 0.0 & 3.000 & 5.196 & 0.192 & unstable \\
			1.0 & 0.5 & 3.081 & 5.641 & 0.177 & unstable \\
			1.0 & 1.0 & 3.295 & 7.126 & 0.140 & unstable \\
			1.2 & 0.5 & 3.500 & 6.861 & 0.146 & unstable \\
			1.5 & 0.5 & 4.129 & 8.665 & 0.115 & unstable \\
			\hline
		\end{tabular}
	\end{table}
	
	\subsection{Marginal stability}
	
	The stability of a circular timelike orbit at fixed $E$ and $L$ is determined by
	the sign of $\mathcal{R}''(r_c)$: the orbit is stable when
	$\mathcal{R}''(r_c)<0$, equivalently $d^2V_{\rm eff}/dr^2|_{r_c}>0$, and
	unstable when the inequalities are reversed. Marginally stable orbits therefore
	satisfy
	\begin{equation}\label{eq.ISCO}
	\mathcal{R}(r_c)=0,\qquad
	\mathcal{R}'(r_c)=0,\qquad
	\mathcal{R}''(r_c)=0.
	\end{equation}
	The unique marginally stable orbit, which we identify as the ISCO, is
	characterised by $\mathcal{R}''(r_c)=0$ with $E$ and $L$ fixed by
	Eqs.~\eqref{eq.Ec}--\eqref{eq.Lc}. In practice this coincides with the extremum
	of $L^2(r)$,
	\begin{equation}
	\frac{dL^2}{dr}=0,
	\end{equation}
	in agreement with condition \eqref{eq.ISCOcond}.
	
	The Schwarzschild limit $(k,\alpha)=(1,0)$ is recovered exactly (in units
	$M=1$):
	\begin{equation}
	r_{\rm ISCO}=6,\quad
	E_{\rm ISCO}=\sqrt{\tfrac{8}{9}},\quad
	L_{\rm ISCO}=2\sqrt{3},\quad
	\Omega_{\rm ISCO}=\frac{1}{6\sqrt{6}},
	\end{equation}
	with the dimensionful frequency $\Omega_{\rm ISCO}=1/(6\sqrt{6}\,M)$.
	For nonzero $k$ and $\alpha$, the ISCO depends on both parameters in a way
	that is sensitive to the choice of normalisation. In seed-mass units, at
	fixed $k=1$ one has $r_{\rm ISCO}/M=6.000$ $(\alpha=0)$, $6.236$
	$(\alpha=0.5)$, $6.779$ $(\alpha=1)$, so increasing $\alpha$ shifts the
	ISCO outward; at fixed $\alpha=0.5$, $r_{\rm ISCO}/M$ increases
	monotonically from $3.133$ at $k=0.5$ through $6.236$ at $k=1$ to $9.054$
	at $k=1.5$. Thus in seed-mass units, increasing $k$ shifts the ISCO
	outward, while decreasing $k$ shifts it inward; the dependence on $\alpha$
	is comparatively mild. Representative values are summarised in
	Table~\ref{tab:marginal_stability}. The sign of $d^2V_{\rm eff}/dr^2$
	changes precisely once across the circular-orbit family in all cases
	examined, confirming a single ISCO with no secondary stable islands.
	
	\begin{table}[t]
		\centering
		\caption{Representative marginally stable circular orbits for the ZVD
			spacetime (units $M=1$, with $M$ the seed length scale). The ISCO
			radius is determined by $dL^2/dr=0$; $E_{\rm ISCO}, L_{\rm ISCO},
			\Omega_{\rm ISCO}$ are evaluated from
			Eqs.~\eqref{eq.Ec}--\eqref{eq.Omega} at $r=r_{\rm ISCO}$. The final
			two columns give the ADM-normalised quantities, with
			$\mathcal{M}_{\rm ADM}=\tfrac{kM}{2}(1+\cosh^2\alpha)$ the physical
			ADM mass.}
		\label{tab:marginal_stability}
		\begin{tabular}{cccccccc}
			\hline
			$k$ & $\alpha$ & $r_{\rm ISCO}$ & $E_{\rm ISCO}$ & $L_{\rm ISCO}$
			& $\Omega_{\rm ISCO}$ & $r_{\rm ISCO}/\mathcal{M}_{\rm ADM}$
			& $\mathcal{M}_{\rm ADM}\Omega_{\rm ISCO}$ \\
			\hline
			0.5 & 0.5 & 3.133 & 0.928 & 1.775 & 0.176 & 5.517 & 0.100 \\
			0.8 & 0.5 & 5.074 & 0.938 & 3.001 & 0.084 & 5.584 & 0.076 \\
			1.0 & 0.0 & 6.000 & 0.943 & 3.464 & 0.068 & 6.000 & 0.068 \\
			1.0 & 0.5 & 6.236 & 0.939 & 3.787 & 0.065 & 5.491 & 0.074 \\
			1.0 & 1.0 & 6.779 & 0.924 & 4.875 & 0.058 & 4.010 & 0.098 \\
			1.2 & 0.5 & 7.372 & 0.940 & 4.568 & 0.053 & 5.409 & 0.072 \\
			1.5 & 0.5 & 9.054 & 0.941 & 5.732 & 0.042 & 5.315 & 0.072 \\
			\hline
		\end{tabular}
	\end{table}
	
\begin{figure}[t]
		\centering
		\includegraphics[width=\textwidth]{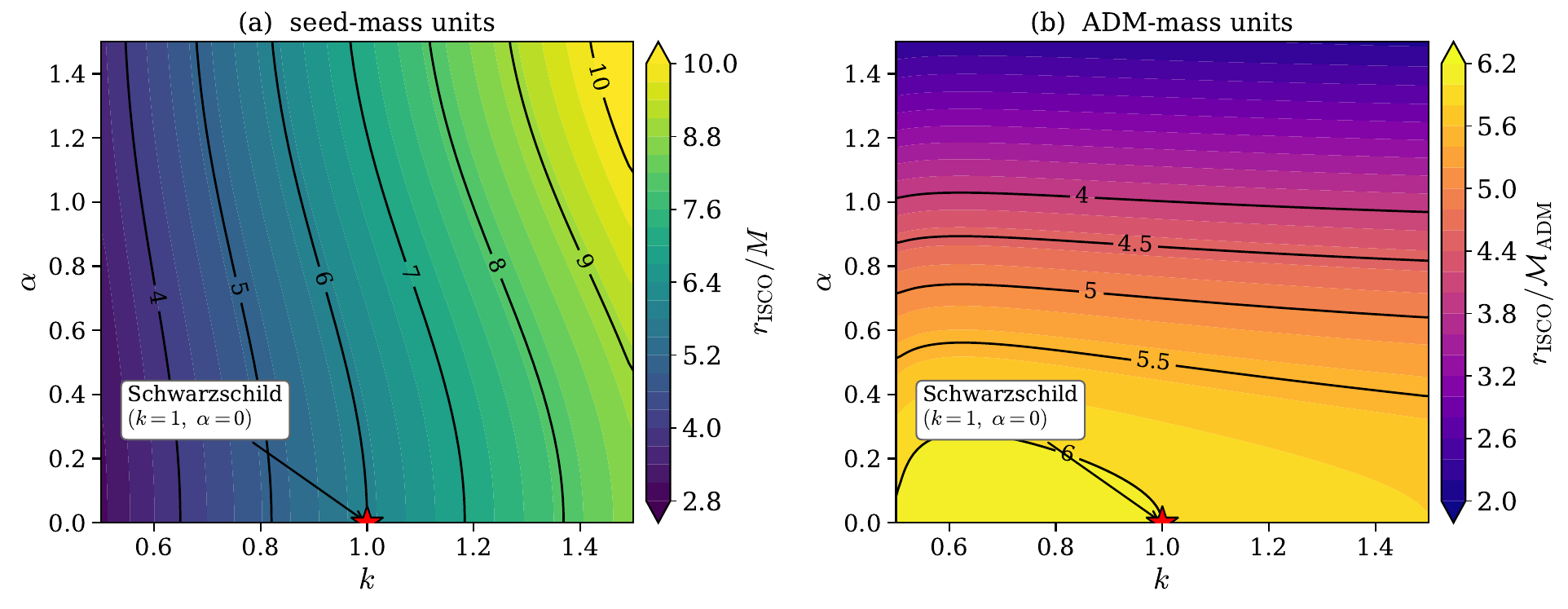}
		\caption{Contour plot of the ISCO radius over the parameter region
			$(k,\alpha)\in[0.5,1.5]\times[0,1.5]$, computed numerically from
			$dL^2/dr=0$. \textbf{(a)} In seed-mass units $r_{\rm ISCO}/M$:
			contours are nearly vertical and the deformation parameter $k$
			drives the variation; increasing either $k$ or $\alpha$ shifts the
			ISCO outward in this normalisation. \textbf{(b)} In ADM-mass units
			$r_{\rm ISCO}/\mathcal{M}_{\rm ADM}$ with
			$\mathcal{M}_{\rm ADM}=\tfrac{kM}{2}(1+\cosh^2\alpha)$: contours
			are nearly horizontal and the boost parameter $\alpha$ drives the
			variation; increasing $\alpha$ moves the ISCO inward, and the ISCO
			falls \emph{inside} the Schwarzschild value $6\mathcal{M}_{\rm ADM}$
			for $\alpha\gtrsim 0.5$ across most of the surveyed range. The red
			star in each panel marks the Schwarzschild limit
			$(k,\alpha)=(1,0)$, where both normalisations agree at
			$r_{\rm ISCO}=6$. The plots terminate on the left at $k=0.5$, on
			the boundary of the region in which a genuine exterior photon
			ring exists; for $k<0.5$ the ISCO continues to exist but the
			associated circular-orbit family extends down to the naked
			singularity at $r=2M$.}
		\label{fig:risco_contour}
	\end{figure}

	A two-dimensional view of the ISCO over the parameter region
	$0.5\le k\le 1.5$, $0\le\alpha\le 1.5$ is shown in
	Figure~\ref{fig:risco_contour}, in both seed-mass and ADM-mass units. The
	two normalisations tell qualitatively different stories. In seed-mass
	units (panel (a)), the contours of constant $r_{\rm ISCO}/M$ are roughly
	vertical: $k$ controls the bulk of the variation, the dependence on
	$\alpha$ is mild, and increasing $k$ from $0.5$ to $1.5$ at fixed
	$\alpha=0.5$ shifts the ISCO from $\sim 3.1M$ to $\sim 9.1M$. In
	ADM-mass units (panel (b)), the contours of constant
	$r_{\rm ISCO}/\mathcal{M}$ are nearly horizontal: $\alpha$ becomes the
	dominant parameter, and the ISCO can lie inside the Schwarzschild value
	$r_{\rm ISCO}=6\mathcal{M}$ once the boost parameter is sufficiently
	large. For example, $(k,\alpha)=(1,1)$ gives $\mathcal{M}\simeq 1.69M$
	and $r_{\rm ISCO}/\mathcal{M}\simeq 4.01$, well inside the Schwarzschild
	value. The trend with $k$ in ADM-mass units is, by contrast, very weak:
	along the $\alpha=0.5$ line, $r_{\rm ISCO}/\mathcal{M}$ varies only
	between $\sim 5.3$ and $\sim 5.6$ as $k$ ranges over the entire surveyed
	interval. This reflects the algebraic identity
	$(\mathcal{M},Q,\Sigma)|_{\rm ZVD}=k(\mathcal{M},Q,\Sigma)|_{k=1}$ noted
	in Section~\ref{sec.properties}: rescaling the seed mass by a factor $k$
	rescales the ADM mass by the same factor, and ratios of geometric
	quantities to $\mathcal{M}$ become approximately $k$-insensitive.
	
	The choice between the two normalisations is physical, not aesthetic.
	Observational comparisons with accretion-disc spectra typically constrain
	$r_{\rm ISCO}/\mathcal{M}$, since $\mathcal{M}$ is the asymptotic mass
	measured by distant observers. In that physically motivated setting,
	panel (b) of Figure~\ref{fig:risco_contour} is the relevant one, and the
	main message is that the Kaluza--Klein dressing acts \emph{inward} on the
	ISCO --- the opposite of the ``outward shift'' suggested by panel (a).
	
	\subsection{Gravitational redshift near the singular surface}
	\label{sec.redshift}
	
	The circular-orbit analysis above is naturally complemented by the
	gravitational redshift, which provides a direct observational diagnostic of
	how deep into the potential the various orbits sit. For a static emitter at
	radius $r$ and a static receiver at infinity, the redshift factor, equivalently
	the emitted-to-received frequency ratio, in the static spacetime
	\eqref{eq.eqmetric} is fixed by the $tt$ component alone,
	\begin{equation}\label{eq.redshift}
	1+z=\left(\frac{-g_{tt}(\infty)}{-g_{tt}(r)}\right)^{1/2}
	=\bigl[A(r)\bigr]^{-1/2}
	=\left(\frac{H(r)}{f(r)^{k}}\right)^{1/2},
	\end{equation}
	where we used $g_{tt}(\infty)=-1$ and $A=f^k/H$. Because the boost enters only
	through the warp factor $H$, which is finite and tends to $\cosh\alpha$ as
	$r\to 2M^{+}$, the redshift is controlled by $f^{k}$: since $f\to0^{+}$ and
	hence $f^{k}\to0$ for every $k>0$, one has $-g_{tt}\to0$ and
	\begin{equation}
	1+z\;\longrightarrow\;\infty\qquad\text{as}\qquad r\to 2M^{+}.
	\end{equation}
	The surface $r=2M$ is therefore an \emph{infinite-redshift surface} for the
	entire ZVD family, including the deformed cases $k\neq1$ where it is a naked
	curvature singularity rather than a regular horizon. The boost raises the
	redshift at fixed $r$ through the factor $H\ge1$ but does not move the
	infinite-redshift surface, which is pinned at $r=2M$ by the seed. In the
	Schwarzschild limit $(k,\alpha)=(1,0)$, Eq.~\eqref{eq.redshift} reduces to the
	familiar $1+z=(1-2M/r)^{-1/2}$.
	
\begin{figure}[t]
		\centering
		\includegraphics[width=0.72\textwidth]{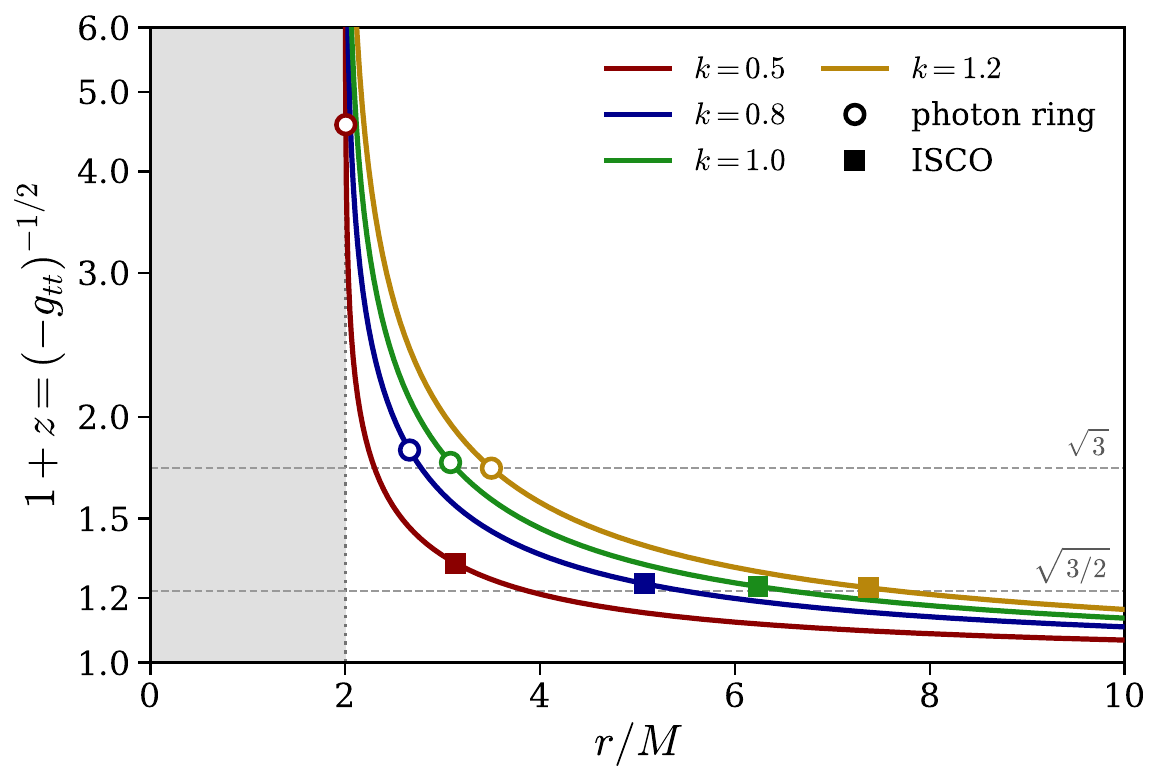}
		\caption{Gravitational redshift factor $1+z=(-g_{tt})^{-1/2}$ for static
			emitters in the ZVD spacetime as a function of the Schwarzschild-like
			radial coordinate $r/M$ (not an areal radius for $k\neq1$), for
			$k=0.5,0.8,1.0,1.2$ at fixed $\alpha=0.5$. All curves diverge at the
			infinite-redshift surface $r=2M$ (dotted line), which is a regular
			horizon only for $k=1$ and a naked curvature singularity otherwise.
			Open circles mark the photon ring $r_{\rm ph}$ and filled squares the
			ISCO $r_{\rm ISCO}$. The clean $k>1/2$ photon rings carry a moderate
			redshift near the Schwarzschild value $\sqrt{3}$, whereas the
			$k=0.5$ ring grazes the singular surface and is strongly redshifted
			($1+z\simeq4.6$); the ISCOs are only mildly redshifted in all cases.
			The vertical axis is logarithmic, which separates the photon-ring and
			ISCO values clearly while still displaying the divergence at $r=2M$;
			the shaded strip $r\le2M$ lies outside the exterior domain, and the
			horizontal dashed lines mark the Schwarzschild reference values
			$1+z=\sqrt{3}$ at the photon ring and $1+z=\sqrt{3/2}$ at the ISCO.}
		\label{fig:redshift}
	\end{figure}

	This behaviour dovetails with the circular-orbit structure of
	Section~\ref{sec.geodesics}. Figure~\ref{fig:redshift} shows
	$1+z(r)$ for several $k$ at $\alpha=0.5$, with the photon ring (open circles)
	and ISCO (filled squares) marked on each curve. For the physically clean
	photon rings with $k>1/2$ the redshift at the ring is moderate and close to
	the Schwarzschild value $1+z=\sqrt{3}\simeq1.732$ (e.g.\ $1.732$ at
	$(k,\alpha)=(1,0)$ and $1.758$ at $(1,0.5)$). As $k\to\tfrac12^{+}$, however,
	the photon ring is driven down towards the singular surface
	(Section~\ref{sec.null}) and its redshift grows without bound: the
	near-singularity ring at $(k,\alpha)=(0.5,0.5)$, with $r_{\rm ph}\simeq2.006M$,
	already has $1+z\simeq4.56$, and for $k<1/2$ the inner edge of the timelike
	family coincides with the infinite-redshift surface itself. Thus the formal
	circular orbits that approach $r=2M$ are not only mathematically extremal in a
	region of unbounded curvature, as emphasised in
	Section~\ref{sec.geodesics}, but also infinitely redshifted: any radiation
	they emit is arbitrarily degraded before reaching a distant observer. This
	reinforces, on observational grounds, the distinction drawn in
	Section~\ref{sec.null} between the physically relevant $k>1/2$ photon ring and
	the near-singularity orbits at $k\lesssim1/2$, which are effectively invisible.
	The ISCO, by contrast, sits well outside the photon ring
	($r_{\rm ISCO}>r_{\rm ph}$) and carries only a mild redshift across the
	surveyed range, from $1+z=\sqrt{3/2}\simeq1.225$ in the Schwarzschild limit to
	$1+z\simeq1.32$ at the most deformed and most strongly boosted points
	(Table~\ref{tab:marginal_stability}); the redshift of the inner edge of an
	accretion disc in the ZVD geometry is therefore comparable to, but somewhat
	larger than, in Schwarzschild.
	
	\subsection{Critical impact parameter and equatorial shadow scale}
	\label{sec.shadow}
	
	The critical impact parameter of the unstable photon ring determines the
	characteristic angular size of the shadow cast by the central object; in the
	Schwarzschild limit this reduces to the standard black-hole shadow radius,
	while for $k\neq1$, where the spacetime carries a naked singularity rather
	than a horizon (Section~\ref{sec.properties}), it should be understood as the
	apparent size of the dark region produced by photon capture toward the
	central singular surface. For equatorial null geodesics
	the impact parameter is $b=L/E$, and the photon ring at $r_{\rm ph}$ fixes the
	critical value \eqref{eq.bph},
	\begin{equation}
	b_{\rm ph}=\left(\frac{C(r_{\rm ph})}{A(r_{\rm ph})}\right)^{1/2},
	\end{equation}
	already tabulated in Table~\ref{tab:null_circular}. In the equatorial sector,
	rays with $b>b_{\rm ph}$ are deflected and escape, while rays with
	$b<b_{\rm ph}$ spiral inward and reach the central singular surface $r=2M$
	(or, in the Schwarzschild limit, the horizon), so for an observer at large
	distance $D$ the apparent angular radius of the dark region is
	$\theta_{\rm sh}\simeq b_{\rm ph}/D$. We therefore take $b_{\rm ph}$ as the
	characteristic shadow radius; in the Schwarzschild limit it reduces to the
	standard $b_{\rm ph}=3\sqrt{3}\,M\simeq5.196M$.
	
	Two caveats should be stated plainly. First, because $k\neq1$ breaks spherical
	symmetry, the shadow of the ZVD spacetime is not in general a perfect circle:
	photons launched out of the equatorial plane probe the $\theta$-dependence of
	the metric \eqref{eq.ZVKKmetric}, and the Hamilton--Jacobi equation does not
	separate for $k\neq1$, so the apparent boundary acquires a mild
	non-circularity. A complete determination of the shadow shape requires
	numerical integration of the off-equatorial null geodesics and is left to the
	ray-tracing study mentioned in Section~\ref{sec.conclusion}. The quantity
	$b_{\rm ph}$ computed here is the exact \emph{equatorial} extent of the shadow
	and sets its characteristic angular scale, which is the relevant quantity for
	a leading-order comparison with horizon-scale images. Second, for $k\le1/2$
	the exterior photon ring degenerates (Section~\ref{sec.null}) and a sharply
	defined shadow edge in the present sense no longer exists; we therefore restrict
	the shadow map to $k>1/2$.
	
\begin{figure}[t]
		\centering
		\includegraphics[width=\textwidth]{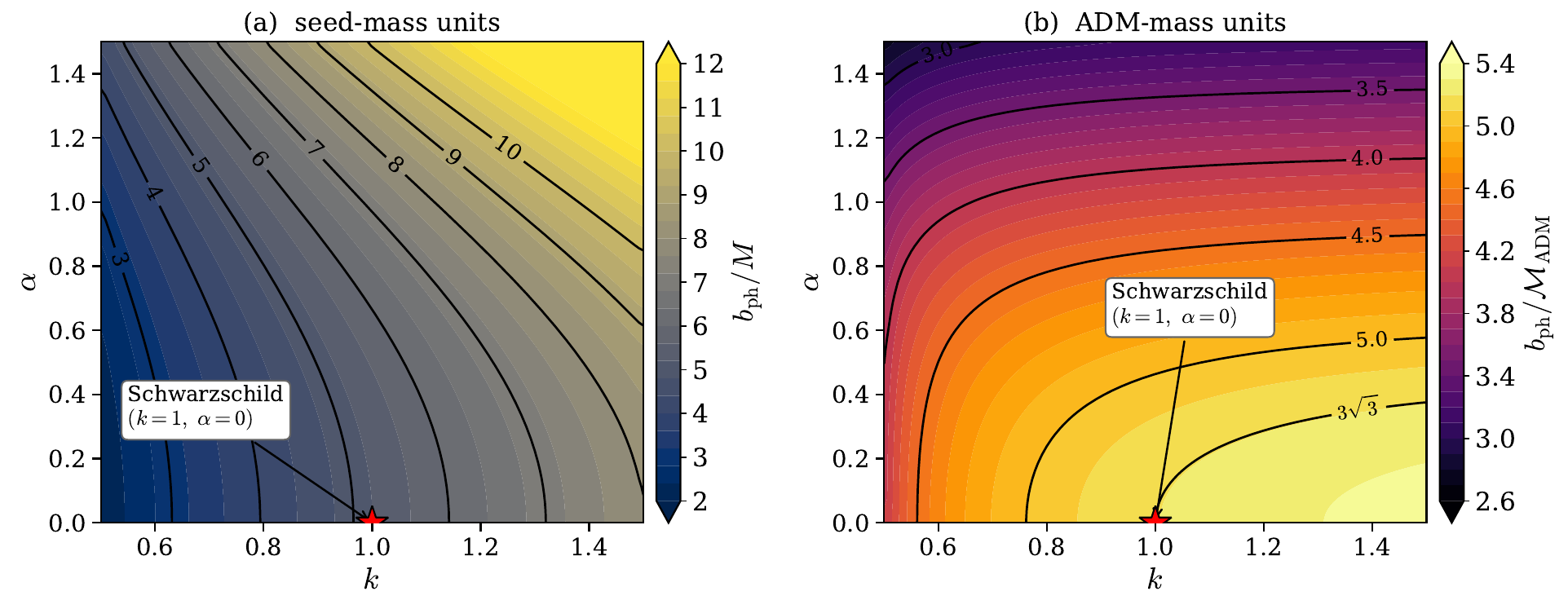}
		\caption{Characteristic shadow radius, i.e.\ the critical photon impact
			parameter $b_{\rm ph}$, over the region
			$(k,\alpha)\in[0.5,1.5]\times[0,1.5]$. \textbf{(a)} In seed-mass units
			$b_{\rm ph}/M$: both $k$ and $\alpha$ enlarge the photon capture
			region. \textbf{(b)} In ADM-mass units
			$b_{\rm ph}/\mathcal{M}_{\rm ADM}$ with
			$\mathcal{M}_{\rm ADM}=\tfrac{kM}{2}(1+\cosh^2\alpha)$: contours are
			nearly horizontal, the boost $\alpha$ drives the variation, and the
			shadow shrinks below the Schwarzschild value $3\sqrt3$ (heavy contour)
			for $\alpha\gtrsim0.5$. The red star marks the Schwarzschild limit
			$(k,\alpha)=(1,0)$, where $b_{\rm ph}=3\sqrt3\,M$ in both
			normalisations. The non-circularity of the true shadow for $k\neq1$ is
			not captured by $b_{\rm ph}$ alone and is left to a future ray-tracing
			analysis; $b_{\rm ph}$ gives the exact equatorial extent.}
		\label{fig:shadow_contour}
	\end{figure}

	Figure~\ref{fig:shadow_contour} maps $b_{\rm ph}$ over the surveyed
	$(k,\alpha)$ region in both seed-mass and ADM-mass units, paralleling the ISCO
	map of Figure~\ref{fig:risco_contour}. The two normalisations again tell
	complementary stories. In seed-mass units [panel (a)] both $k$ and $\alpha$
	enlarge $b_{\rm ph}/M$: the photon capture cross-section grows with the
	deformation and with the Kaluza--Klein dressing, consistent with the
	photon-ring radii of Table~\ref{tab:null_circular}. In ADM-mass units
	[panel (b)] the picture inverts, exactly as for the ISCO:
	$b_{\rm ph}/\mathcal{M}_{\rm ADM}$ \emph{decreases} with $\alpha$ and falls
	below the Schwarzschild value $3\sqrt3$ once the boost is appreciable. For
	example, at $(k,\alpha)=(1,1)$ one has $\mathcal{M}_{\rm ADM}\simeq1.69M$ and
	$b_{\rm ph}/\mathcal{M}_{\rm ADM}\simeq4.2$, about $19\%$ below the
	Schwarzschild $3\sqrt3\simeq5.196$. By the same $k$-rescaling identity
	$(\mathcal{M},Q,\Sigma)|_{\rm ZVD}=k\,(\mathcal{M},Q,\Sigma)|_{k=1}$ of
	Section~\ref{sec.properties}, the deformation parameter drops out of
	$b_{\rm ph}/\mathcal{M}_{\rm ADM}$ almost entirely, leaving $\alpha$ as the
	controlling variable. The physical reading mirrors that of
	Section~\ref{sec.null}: at fixed asymptotic mass, the electric/dilaton
	dressing \emph{shrinks} the shadow, so that in physical mass units the
	imprint of the Kaluza--Klein charge on a horizon-scale image is a reduction,
	not an enlargement, of the shadow size.
	
	\section{Conclusion}
	\label{sec.conclusion}
	
	We have derived and analysed a new exact solution of four-dimensional
	Kaluza--Klein theory, which we call the Zipoy--Voorhees--dilaton (ZVD)
	spacetime. The solution was obtained by applying the uplift--boost--reduction
	procedure to the Zipoy--Voorhees vacuum seed, thereby endowing the static
	axisymmetric deformation of the Schwarzschild geometry with an electric charge
	and a nontrivial dilaton profile. The spacetime is parametrised by the seed
	length scale $M$, the Zipoy--Voorhees deformation parameter $k$, and the
	boost parameter $\alpha$, and is given explicitly in
	Eqs.~\eqref{eq.ZVKKmetric}--\eqref{eq.PhiZV}.
	
	The asymptotic conserved charges are
	\[
	\mathcal{M}=\frac{kM}{2}(1+\cosh^2\alpha),\quad
	Q=\frac{kM}{2}\sinh 2\alpha,\quad
	\Sigma=\frac{\sqrt{3}}{2}\,kM\sinh^2\alpha,\quad J=0,
	\]
	and reduce to the expected values in the limits $k=1$ and $\alpha=0$. The
	dilaton charge satisfies the constraint $\Sigma^2+\sqrt{3}\,\mathcal{M}\Sigma
	=\frac{3}{2}Q^2$ for all $k$ and $\alpha$, generalising the relation found in
	\cite{Aliev:2008wv} and showing that $\Sigma$ is determined by $\mathcal{M}$
	and $Q$ irrespective of the quadrupole deformation. The ZVD spacetime is
	asymptotically flat in the standard sense for every $k$, with the
	elementary-flatness ratio of \eqref{eq.flatratio} approaching unity on the
	half-axis away from the singular surface. For $k=1$ it admits a regular
	black-hole interpretation; for $k\neq 1$ the surface $r=2M$ is a naked
	curvature singularity inherited from the vacuum Zipoy--Voorhees seed, and
	the black-hole interpretation does not apply.
	
	The equatorial geodesic structure was analysed in detail for neutral test
	particles and photons. The null circular orbit and the inner edge of the
	timelike circular family are both governed by the condition $D(r)=0$ for
	the explicit elementary function $D(r)=2[r-1-\Upsilon(r)]$. The behaviour
	at the singular surface, $D(2M^+)=2(1-2k)$, together with numerical root
	counting over the surveyed parameter region, locates a unique exterior
	photon-ring threshold at $k=1/2$. This generalises the known
	$\gamma>1/2$ threshold of the vacuum Zipoy--Voorhees metric
	\cite{Kodama:2003ch} to the electrically charged Kaluza--Klein case, with
	the boost parameter $\alpha$ shifting the photon-ring radius outward
	without altering the threshold itself. When present, the photon ring is
	always unstable. For timelike orbits, the surveyed parameter range reveals
	a single connected family with a unique ISCO. The dependence on $(k,\alpha)$
	is normalisation-sensitive: in seed-mass units (panel (a) of
	Figure~\ref{fig:risco_contour}), increasing $k$ shifts $r_{\rm ISCO}/M$
	outward and increasing $\alpha$ shifts it outward more mildly; in
	ADM-mass units (panel (b)), $\alpha$ becomes the dominant variable and
	pushes $r_{\rm ISCO}/\mathcal{M}$ inward, with the ISCO falling inside
	the Schwarzschild value $6\mathcal{M}$ for $\alpha\gtrsim 0.5$ across
	most of the surveyed range. For observational comparisons in physical
	mass units the second normalisation is the relevant one.
	
	Several directions suggest themselves for future work. First, a full
	ray-tracing computation of the optical shadow --- building on the
	photon-ring radii and critical impact parameters determined here --- could
	in principle test the imprint of the dilaton/electric dressing on shadow
	morphology, although direct comparison with Event Horizon Telescope targets
	\cite{EventHorizonTelescope:2019dse,EventHorizonTelescope:2022xnr,
		EventHorizonTelescope:2022vjs} would require the rotating extension
	below. Second, the most astrophysically relevant extension is to a
	rotating configuration, and the uplift--boost--reduction procedure used
	here is not tied to a static seed. Applied to a stationary, axisymmetric
	vacuum seed it produces a four-dimensional Kaluza--Klein electric/dilaton
	spacetime carrying both angular momentum and charge. Explicit rotating
	generalisations of the Zipoy--Voorhees metric are already available and
	provide natural candidate seeds. The stationary version of the ZV/$q$-metric,
	whose relativistic multipole moments have been computed by means of the
	Fodor--Hoenselaers--Perj\'es formalism \cite{FrutosAlfaro:2018}, and the
	$\delta$-Kerr spacetime, obtained as a nonlinear superposition of the
	$\delta$-metric (the ZV metric with $\delta=k$) and the Kerr solution and
	whose quasinormal-mode spectrum has been determined in both the light-ring
	and wave approaches \cite{Allahyari:2020}, are the most directly relevant;
	the Tomimatsu--Sato family and, more generally, the rotating Weyl solutions
	generated from the ZV seed by the Ernst or
	Hoenselaers--Kinnersley--Xanthopoulos techniques provide further options.
	Taking such a seed as $\tilde g_{\mu\nu}$ and repeating the five-dimensional
	boost would yield a ``rotating ZVD'' spacetime. It is worth noting that for
	the $\delta$-Kerr geometry the outer singularity remains a closed null
	hypersurface over a finite range of the deformation parameter
	\cite{Allahyari:2020}, so that rotation can restore a one-way outer boundary
	of the kind that is absent from the static geometry for $k\neq1$ --- an
	appealing feature for the charged extension envisaged here. Two features distinguish this from the
	static case treated here. First, for a stationary seed the boost no longer
	leaves the metric static: the general reduction \eqref{eq.Agen} then
	generates a magnetic component $A_\phi$ alongside $A_t$, the reduced metric
	acquires a $g_{t\phi}$ term, and the geodesic problem loses the
	simplifications that followed from staticity --- in particular the clean
	identity $A(r)C(r)=r(r-2)$ and the elementary form of $D(r)$ that organised
	the whole of Section~\ref{sec.geodesics}. Second, every available rotating
	seed is algebraically far more involved than the static ZV metric: the
	Tomimatsu--Sato potentials are rational rather than power-law in
	prolate-spheroidal coordinates, and the $\delta$-Kerr geometry is in practice
	handled perturbatively in the quadrupole and rotation parameters
	\cite{Allahyari:2020}. The warp factor $H$ and the reduced fields therefore
	become unwieldy, and the photon-ring and ISCO analysis would have to proceed
	largely numerically. The
	interplay of frame dragging, quadrupole deformation, and Kaluza--Klein charge
	in such a spacetime --- and in particular whether the clean $k=1/2$
	photon-ring threshold found here survives in deformed form --- is a natural
	and worthwhile target, and recent rotating Kaluza--Klein constructions with
	shadow and accretion-disc analyses \cite{Hosseinifar:2026} indicate the kind
	of phenomenology such a solution would open up. Third, it would be interesting to
	examine whether a consistent thermodynamic first law can be formulated
	using the ADM charges derived here, despite the absence of a regular
	horizon for $k\neq1$. The ZVD spacetime is best regarded as a controlled
	analytic laboratory for studying how Kaluza--Klein electric/dilaton
	dressing modifies the geodesic and null-orbit diagnostics of a static
	non-Schwarzschild compact object, complementing other charged ZV
	constructions in the literature \cite{Yunusov:2025chw} that arise in
	different theories with different scalar couplings.
	
	\section*{Acknowledgement}
	
	This work was supported by LPPM-UNPAR through the Penelitian Publikasi
	Internasional Bereputasi funding scheme.
	
	\section*{Declaration on the use of generative AI}
	During the preparation of this work the author used Anthropic's Claude
	to assist with language editing and code review. After using this tool,
	the author reviewed and edited the content as needed and takes full
	responsibility for the content of the publication.

\end{document}